\documentstyle{article}
\newcommand{\beq}{\begin{equation}}
\newcommand{\eeq}{\end{equation}}
\newcommand{\beqa}{\begin{eqnarray}}
\newcommand{\eeqa}{\end{eqnarray}}
\newcommand{\bea}{\begin{eqnarray}}
\newcommand{\eea}{\end{eqnarray}}

\newcommand   {\bl}      {\sigma^{(s)}}
\newcommand   {\fl}      {\psi^{(s)}}
\newcommand   {\ft}      {\tilde f_k^2}

\newcommand   {\de}      {(N_+-N_-)}

\newcommand   {\ev}[1]   {\langle #1\rangle}
\newcommand   {\evf}[1]  {\langle #1\rangle_{N,N_+}}

\input{psfig}

\begin{document}
\begin{titlepage}
\begin{flushright}
LU TP 95-31\\
Revised Version\\
\today \\
\end{flushright}
\vspace{0.2in}
\LARGE
\begin{center}
{\bf Evidence for Non-Random Hydrophobicity Structures in Protein Chains}\\
\vspace{.3in}
\large
Anders Irb\"ack\footnote{irback@thep.lu.se},
Carsten Peterson\footnote{carsten@thep.lu.se},
and 
Frank Potthast\footnote{frank@thep.lu.se}\\
\vspace{0.05in}
Department of Theoretical Physics, University of Lund \\ 
S\"{o}lvegatan 14A,  S-223 62 Lund, Sweden \\
\vspace{0.3in}

Published in \\
 {\it Proceedings of the National Academy of Sciences USA }\\
 {\bf 93}, 9533-9538 (1996)\\

\end{center}
\vspace{0.1in}
\normalsize

Abstract:

The question of whether proteins originate from random sequences of amino 
acids is addressed. A statistical analysis is performed in terms of blocked 
and random walk values formed by binary hydrophobic assignments of the 
amino acids along the protein chains. Theoretical expectations of these variables 
from random distributions of hydrophobicities are compared with those obtained  
from functional proteins. The results, which are based upon  proteins in the  SWISS-PROT data base,  convincingly  show that the amino acid sequences 
in proteins differ from what is expected from random sequences in a statistically  significant way. By performing Fourier transforms on the random walks one 
obtains additional evidence for non-randomness of the distributions. 

We have also analyzed results from a synthetic model containing only two amino-acid 
types, hydrophobic and hydrophilic. With reasonable criteria on good folding 
properties in terms of thermodynamical and kinetic behavior, sequences that fold 
well are isolated. Performing the same statistical analysis on the sequences that 
fold well indicates similar deviations from randomness as for the functional 
proteins. The deviations from randomness can be interpreted as originating 
from anticorrelations in terms of an Ising spin model for the hydrophobicities. 

Our results, which differ from some previous investigations using other methods, 
might have impact on how permissive with respect to sequence specificity 
the protein folding process is -- only sequences with non-random hydrophobicity 
distributions fold well.  Other distributions give rise to energy landscapes with 
poor folding properties and hence did not survive the evolution.

\end{titlepage}

\newpage

\section{Introduction}

Hydrophobicity is widely believed to play a central role in the formation of 3D 
protein structures. To understand the statistical distribution of hydrophobicity
along proteins is therefore of utmost interest. This question has been
addressed previously. In Ref.~\cite{white} the authors used binary hydrophobicity 
assignments, zero or one, and studied simultaneously the distribution of clumps
of both zeros and ones by using the so-called run test. For the majority of the 
proteins examined it was found that the results could not be distinguished from 
those corresponding to completely random sequences. The same type of 
statistical test has also been applied to sequences stemming from a simplified 
protein model~\cite{shakhnovich}. Here randomly selected sequences were compared 
with sequences that had been specially designed to have good folding properties. 
The statistical analysis did not reveal any difference between these two groups. 
These findings  seem to indicate that the folding requirements on proteins are 
fairly permissive with little sequence specificity. A slightly different approach 
to analyze the same problem was pursued in Ref.~\cite{pande}, where 
by mapping the binary chains  onto the trajectories of a random walk, 
deviations from random distributions are reported.

Also,  recent work on simplified models suggest the non-randomness 
\cite{sali,irback}. In these studies a large number of randomly 
selected sequences were investigated,and it was found that only a small 
fraction of them folded easily into a thermodynamically stable state. 

In this work we study the statistical distribution of hydrophobicity by using 
methods different from the run test in Ref.~\cite{white}. Along the same lines 
as in Ref.~\cite{pande} rather than analyzing raw sequences of hydrophobicity, 
we focus on the corresponding random walk representation. In this way 
the analysis is more sensitive to long range correlations along the sequence. 
Our analysis has been carried out using two different methods, which differ 
substantially from what is used in Ref.~\cite{pande} although the starting 
point is similar.
First, we form block variables, and study how the behavior of these 
depends on the block size. When applied to the SWISS-PROT data 
base \cite{swiss-prot} of functional proteins, this method yields  
clear evidence for non-randomness. In addition, we have performed a Fourier 
analysis based on the random walk representation. In this analysis 
we find non-random behavior at the wave-length corresponding to 
$\alpha$-helix structure, as one might have expected, but also at 
large wave-lengths.

In our analysis we have divided the sequences into groups corresponding 
to different fractions of hydrophobic residues. This division is important
because the results for different groups deviate in different directions 
from those for random sequences. For sequences with a typical fraction      
of hydrophobic residues we find that the non-randomness can be
interpreted as anticorrelations. This interpretation emerges from  a
simple Ising model of antiferromagnetic interactions among the residues.

Given the impact our results might have on the issue of how permissive with 
respect to sequence specificity the protein folding process is, we have 
carried out 
the same analysis for a toy model~\cite{still1,still2}, for which unbiased
samples of folding and non-folding sequences can be obtained. 
This model, hereafter denoted the {\bf AB} model, consists of chains 
of two kinds of ``amino acids'' interacting with Lennard-Jones potentials. 
We have examined the behavior of 300 randomly selected chains of length 20 
in this model~\cite{irback1}. Out of these only about 10\%  were found to 
have reasonable folding properties. Analyzing these sequences with 
the same methods as being used for the functional proteins, we obtain
results that are qualitatively very similar to those for proteins with a
typical fraction of hydrophobic residues. In particular, we again find 
deviations from random behavior that correspond to anticorrelations.  
One should keep in mind that the toy model chains are quite 
short and highly simplified as compared to functional proteins. 
Nevertheless, it is appealing to attempt an explanation for the 
observed similarity in behavior as originating from the fact that  
those amino acid sequences exhibiting this type of hydrophobicity 
distribution are the ones that fold well. Other distributions give rise 
to energy landscapes with poor folding properties and hence did 
not survive the evolution.

All our analysis concerns comparisons between distributions. The ultimate 
challenge is to decide whether a given sequence is non-random or not. This 
issue, which is beyond the scope of the paper, may be feasible when combining 
different cuts on the measures developed here.

This paper is organized as follows. In Section 2 we develop our two methods 
for analyzing binary hydrophobicity sequences. In Section 3 and 4 these 
methods are applied to real and toy model proteins respectively. Section 4 
also contains the interpretation of deviations from randomness in terms of 
anticorrelations. Finally, a brief summary and outlook can be found in 
Section 5.

\section{Methods}

In this section we describe the variables and statistical methods employed.  
Two different, but not completely 
unrelated approaches are used -- the blocking and Fourier transform methods.  
These will be described in some detail below.

Throughout the paper we consider  sequences of $N$ residues and denote by $\sigma_i$ 
the hydrophobicity of residue $i$. We use a binary hydrophobicity scale: 
$\sigma_i=1$ if residue $i$ is hydrophobic and $\sigma_i=-1$ otherwise. The analysis 
can easily be extended to an arbitrary number of allowed hydrophobicity values and 
we do not expect our results to be affected by using such 
multivalued hydrophobicity  assignments. 

Hydrophobicities $\sigma_i$ represent local properties of a chain. As in Ref.~\cite{pande} 
in order to capture some long range correlation properties we  
consider a random walk representation  
\beq
\label{rw}
r_n=\sum_{i=1}^n\sigma_i
\eeq
for $i=1,\ldots,N$ and where $r_0=0$.

\subsection{The Blocking Method}

Analyzing the behavior of block variables is a widely used and fruitful technique 
in statistical mechanics, and our application will turn out to be no exception. 
For a block size $s$ we define the variables 
\beq 
\bl_i=\sum_{j=1}^s\sigma_{(i-1)s+j}=r_{is}-r_{(i-1)s}
\qquad i=1,\ldots,N/s 
\label{block}
\eeq
where it is assumed that  $N$ is a multiple of $s$. The scaling behavior of  $\bl_i$ 
with increasing $s$ is determined by the correlations between  $\sigma_i$ and $\sigma_j$. 
If the $\sigma_i$'s are independent random numbers drawn from the same distribution,   the correlations between different $\sigma_i$ and  $\sigma_j$ vanish and the variance 
of $\bl_i$ scales linearly with $s$.  

We need to be able to compare real proteins  with a random distribution of  
hydrophobic residues. For this reason we average over all sequences with a fixed 
length and composition. These averages are denoted by $\ev{\cdot}_{N,N_+}$, 
where $N$ is the total number of residues and $N_+$ is the number of 
hydrophobic residues. 

In order to study the fluctuations of the block variables we introduce the
normalized variables
\beq
\fl_i={1\over K}
\Bigl(\bl_i-{s\over N}\sum_{j=1}^{N/s}\bl_j\Bigr)^2\qquad i=1,\ldots,N/s
\label{fluct}
\eeq
where
\beq
K={4N_+(N-N_+)\over N(N-1)}(1-s/N)
\label{const}
\eeq
The constant $K$ is chosen such that $\evf{\fl_i}=s$
for all $N$ and $N_+$. The fact that $K$ depends on $s$ implies
that the variance of $\bl_i$ is not linear in $s$, which is due to the
fact that the average is taken at fixed composition. At fixed $s$
this deviation from linearity disappears in the limit $N\to\infty$. 
If all the residues are of the same type, $K$ vanishes. 
Such sequences are uninteresting in the present analysis and have 
therefore been excluded.    

An important quantity is the (normalized) mean-square fluctuation 
of the block variables, defined by
\beq
\fl={1\over K}{s\over N}
\sum_{i=1}^{N/s}\Bigl(\bl_i-{s\over N}\sum_{j=1}^{N/s}\bl_j\Bigr)^2=
{s\over N}\sum_{i=1}^{N/s}\fl_i
\label{msfluct}
\eeq
Obviously, one has 
\beq
\evf{\fl}=s
\label{lin}\eeq
and that  $\psi^{(1)}=1$ is independent of configuration $\sigma_i$. 
It is also important 
to know the variance of $\fl$. The complete expression for this quantity is
lengthy and can be found in  Appendix A. However, in the $N\to\infty$ limit it 
takes the simple form 
\beq
\evf{\fl{}^2}-\evf{\fl}^2\sim {2s^2(s-1)\over N}
\label{varmsfluct}
\eeq

When studying proteins from the data base, we  average over
sequences with different length and composition. For a general
quantity this requires some assumption about the probability
of different values of $N$ and $N_+$ in order to compare with
random sequences. This problem is absent for $\fl$, since it
is defined such that $\evf{\fl}$ is independent of $N$ and $N_+$.
The variance of $\fl$, on the other hand, does depend upon $N$ and $N_+$. 
However, for an interval $N_1<N<N_2$ with both $N_1$ and $N_2$ large and 
$(N_2-N_1)$ not too large, it is still possible to use Eq.~(\ref{varmsfluct})
for estimating  the variance. 

\subsection{The Fourier Transform Method}

The most direct way to detect periodicity in the distribution of 
hydrophobic residues is to use Fourier analysis. It is well-known that
the Fourier component corresponding to a period of 3.6 residues 
tends to be strong for sequences that form $\alpha$ helices
\cite{eisenberg}. Also, sequences that form $\beta$ sheets tend to
exhibit a periodicity in the hydrophobicity of about 2.3 residues.   
In this paper we compare the full power spectrum for proteins with that
for random sequences.  

As a starting point for our Fourier analysis we take the random
walk representation $r_n$. Since we want to compare with
random sequences and since any permutation of the residues leaves 
the end point $r_N$ unchanged, it is here convenient to introduce 
the modified random walk (see Appendix B) 
\bea
\rho_0&=&r_0=0\\
\rho_n&=&\sum_{i=1}^n\Bigl(\sigma_i-{2N_+-N\over N}\Bigr)=
r_n-n{2N_+-N\over N}
\qquad n=1,\ldots,N
\label{modwalk}
\eea
which is defined such that $\rho_0=\rho_N=0$. With these boundary
conditions, we consider the sine transform
\beq
f_k=\sum_{n=1}^{N-1}\rho_n\sin{\pi kn\over N}\qquad k=1,\ldots,N-1
\label{sinetransf}
\eeq
where the $k$th component corresponds to a wave-length of $2N/k$ residues. 

It is easy to see that the average of $f_k$ over all sequences with a 
fixed length and composition vanishes, and for the squared amplitude we find 
\beq    
\evf{f_k^2}={2N_+(N-N_+)\over N-1}{1\over (2\sin {\pi k\over 2N})^2}
\label{fk2}
\eeq
which shows that this quantity behaves as $k^{-2}$ for small $k$. 
In Appendix B we also give the fourth moment of the $f_k$ distribution.

In our calculations we have used the normalized squared amplitude
\beq
\ft=\frac{f_k^2}{\evf{f_k^2}}
\eeq
which has an average $\evf{\ft}=1$  independent of $N$ and $N_+$. 
By measuring $\ft$ one can, of course, only study the relative strength of 
the different components. In fact, it can easily be shown that
\beq
 \sum_{k=1}^{N-1}\ft=N-1
\eeq
independent of configuration $\sigma_i$.

\section{Real Proteins}

Our analysis has been carried out using the SWISS-PROT data base,
release 31 \cite{swiss-prot}. Some proteins were removed from this 
data base due to uncertainties (see Appendix C for details). Also, we limit the 
analysis to proteins with $N\geq 50$ after the endpoints have been removed 
according to a prescription to be dealt with below. To each residue 
we assigned a binary hydrophobicity value, which was taken to be $+1$ for Leu, 
Ile, Val, Phe, Met, and Trp and $-1$ for the others. This choice was done by 
picking the residues with strongest hydrophobic interactions down to a preset level.  Alternative  definitions, with $4$ to $11$\footnote{This is the choice of Ref.~\cite{white}.} of the residues classified as hydrophobic, have also been tested, with qualitatively similar results.

Estimates of statistical errors on the measurements have been obtained by dividing 
the data into 20 groups and treating the corresponding averages as independent measurements. All statistical errors quoted are in $\sigma$ error units.

Before starting our final analysis of hydrophobicity correlations, 
we need to deal with  two important observations. 
\begin{itemize}
\item The data originating from the ends of the sequences display a different
 behavior than the data from the rest of the sequences. 
\item Sequences with different fractions of hydrophobic residues tend to 
behave in different ways. 
\end{itemize}
As a result, important effects can easily be missed out if averages are computed 
over the full data set, as will be seen below.

\subsection{The Interior versus the Ends}     

We begin by examining whether the behavior of the block variable
depends upon the position of the block along the sequence. This analysis is carried 
out  for block sizes $s=2,3,4,6$, and $12$. In order to obtain a sequence that
can be divided into blocks for each of these sizes, we disregard up to eleven 
residues at the ends. In this way we form a sequence of length 
$N^\prime=n\cdot12$, where $n$ is the largest integer such that 
$n\cdot 12\le N$, $N$ being the length of the original sequence.

\begin{figure}[tbp]
\begin{center}
\vspace{-45mm}
\mbox{\hspace{0mm}\psfig{figure=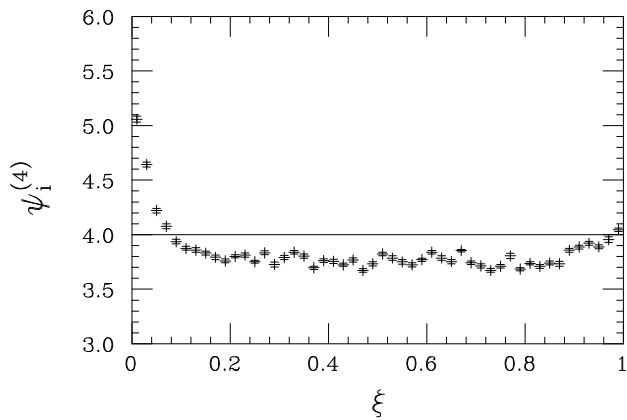,width=10.5cm,height=14cm}}
\vspace{-45mm}
\end{center}
\caption{Average values of $\psi^{(4)}_i$ against the relative positions
of the blocks along the sequence, $\xi$.}  
\label{fig:1}
\end{figure}

We study the block fluctuation $\fl_i$ as a function of the relative 
position $\xi$ of the block center, $\xi$ being 0 at the $N$-end and 1 at 
the $C$-end. The interval in $\xi$ from 0 to 1 is divided into 50 bins 
and average values were computed for each of these bins, using all 
sequences in the data base with more than $50$ residues. 
In Fig.~\ref{fig:1} we show the result for block size $s=4$.
The results for other values of $s$ are similar.
The horizontal line in the figure represents random sequences; 
if the distribution of hydrophobic residues were random, 
the average of $\fl_i$ would be $s$, independent of $i$. 

From Fig.~\ref{fig:1} it is clear that the block fluctuations are roughly 
constant over a wide range in $\xi$. However, it is also evident that the 
fluctuations tend to increase in strength at the ends, in particular at 
the $N$-end. One also notices that the deviations from the random value 
tend to cancel if one averages over all positions.  

This shows that it is important to distinguish between the ends and the 
interior of the sequences when studying hydrophobicity correlations. 
In what follows we focus on the interior by ignoring  15\% of the 
residues at each of the two ends, and analyze  sequences containing 
the remaining 70\% of the residues.                   

\subsection{The Fraction of Hydrophobic Residues}

Our main focus in this paper is on the distribution of hydrophobic  
residues along the sequence, and to what extent this distribution has  
random characteristics. One may also ask whether the total number of 
hydrophobic residues in a sequence follows a random pattern. This question 
can be addressed by studying the quantity  
\beq
X={N_+-Np\over \sqrt{Np(1-p)}}
\label{X}
\eeq
where $N_+$ is the number of hydrophobic residues, $N$ is the
total number of residues, and $p$ is the average of $N_+/N$ over
all sequences. If $N$ hydrophobicity values are drawn randomly and 
independently with probability $p$ for the value 1 and $(1-p)$ for -1, 
the distribution of $X$ becomes approximately Gaussian with zero 
mean and unit variance for large $N$.    

\begin{figure}[tbp]
\begin{center}
\vspace{-45mm}
\mbox{\hspace{0mm}\psfig{figure=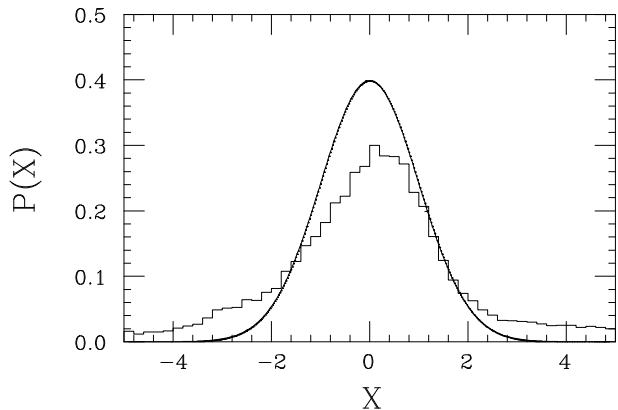,width=10.5cm,height=14cm}}
\vspace{-45mm}
\end{center}
\caption{The distribution of $X$ for the sequences in the data base.  
The curve is the Gaussian with zero mean and unit variance.}  
\label{fig:2}
\end{figure}

We have calculated $X$ for the sequences in the data base, after 
eliminating 30\% of the residues, as discussed above. 
The average fraction of hydrophobic residues 
was found to be $p\approx 0.291$. The distribution of 
$X$ obtained 
is shown in Fig.~\ref{fig:2}, from which we see that the tails are larger 
than for the random distribution. 

When studying correlations in hydrophobicity, we have divided the 
sequences into groups corresponding to different regions in $X$.
This division need not have a simple interpretation in terms of 
standard groups of proteins, but turns out to be useful. Indeed, we
will find below that sequences with different $X$ tend to display 
different types of correlations.

\subsection{Results}

We now turn to the results of our block and Fourier analysis.
As discussed in the previous two subsections, we have chosen to consider 
the interior of the sequences and to study different regions in $X$. 

\begin{figure}[tbp]
\begin{center}
\vspace{-45mm}
\mbox{\hspace{0mm}\psfig{figure=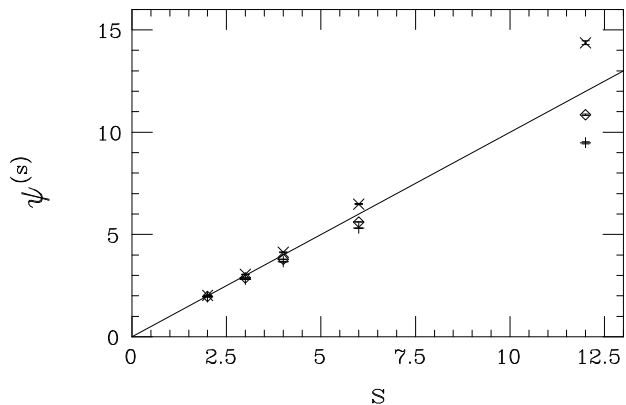,width=10.5cm,height=14cm}}
\vspace{-45mm}
\end{center}
\caption{Mean-square fluctuation of the block variables, $\fl$, as 
a function of block size $s$ for $|X|<0.5$ (+ ;10154 qualifying proteins), $|X|>3$ ($\times$; 4928), and all $X$ ($\diamond$; 36765). The averages have been 
computed over sequences which contained more than 50 residues prior to the 
elimination of residues at the ends. The straight line is the result for random 
sequences.} 
\label{fig:3}
\end{figure}

First we consider the mean-square fluctuation of the block variables, $\fl$. 
In Fig.~\ref{fig:3} results are shown corresponding to three different 
regions in $X$:  $|X|<0.5$, $|X|>3$, and all $X$. The straight 
line represents random sequences. We see that the results for large $X$ lie above 
this line, while the results for small $X$ show the opposite 
behavior. The same pattern is observed when using alternative hydrophobicity 
assignments. Notice that $\fl$ cannot increase slower than 
linearly with $s$ if the correlation between $\sigma_i$ and $\sigma_j$ is 
translationally invariant and non-negative.  Therefore, these results suggest 
that there exists negative hydrophobicity correlations for small $X$.   

\begin{figure}[tbp]
\begin{center}
\vspace{-45mm}
\mbox{\hspace{0mm}\psfig{figure=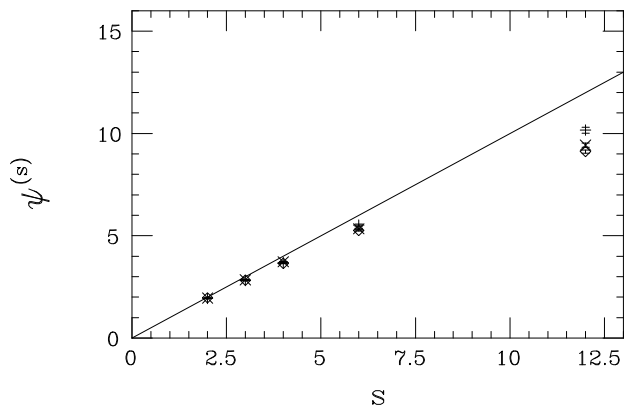,width=10.5cm,height=14cm}}
\vspace{-45mm}
\end{center}
\caption{Mean-square fluctuation of the block variables, $\fl$, against block 
size $s$ for $|X|<0.5$ and $50<N\le 150$ (+; 2457 qualifying proteins), 
$150<N\le 250$ ($\times$; 2228), and $250<N\le 350$ ($\diamond$; 1642). 
The straight line is the result for random sequences.}
\label{fig:4}
\end{figure}

We have also tested how these results depend on the length of the
sequences by computing averages of $\fl$ corresponding to 
different intervals in $N$, where $N$ is the length of the sequence prior to  
the elimination of residues at the ends. In Fig.~\ref{fig:4} we show 
results obtained for $|X|<0.5$ and three different intervals in $N$. 
It is clear from Fig.~\ref{fig:4} that the size dependence is fairly weak. 
Another interesting feature is that the deviation from the result for 
random sequences grows with  sequence length. Notice that 
the variance of $\fl$ scales as $N^{-1/2}$ for random sequences.

\begin{figure}[tbp]
\begin{center}
\vspace{-43mm}
\mbox{\hspace{-31mm}\psfig{figure=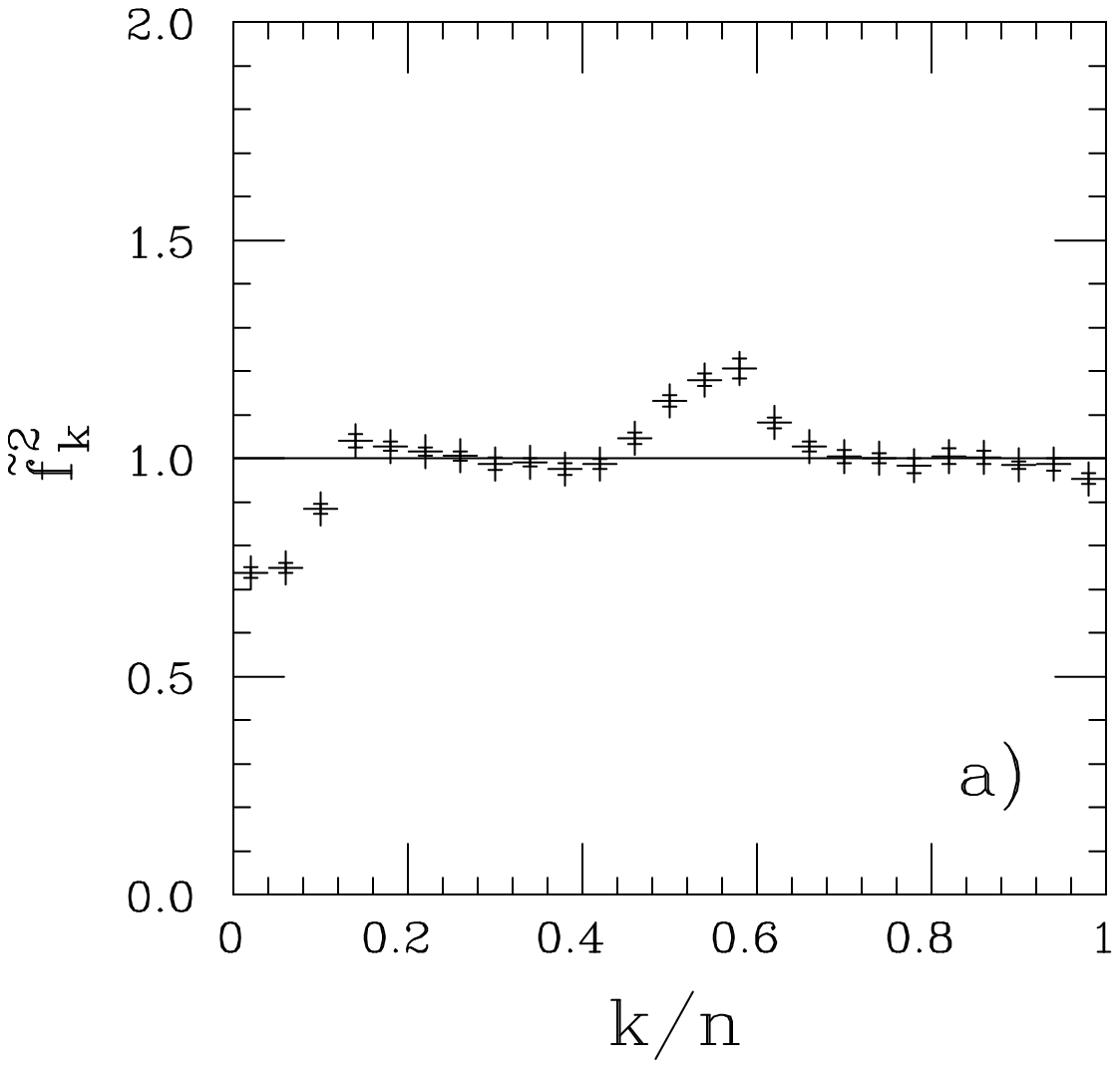,width=10.5cm,height=14cm}
\hspace{-30mm}\psfig{figure=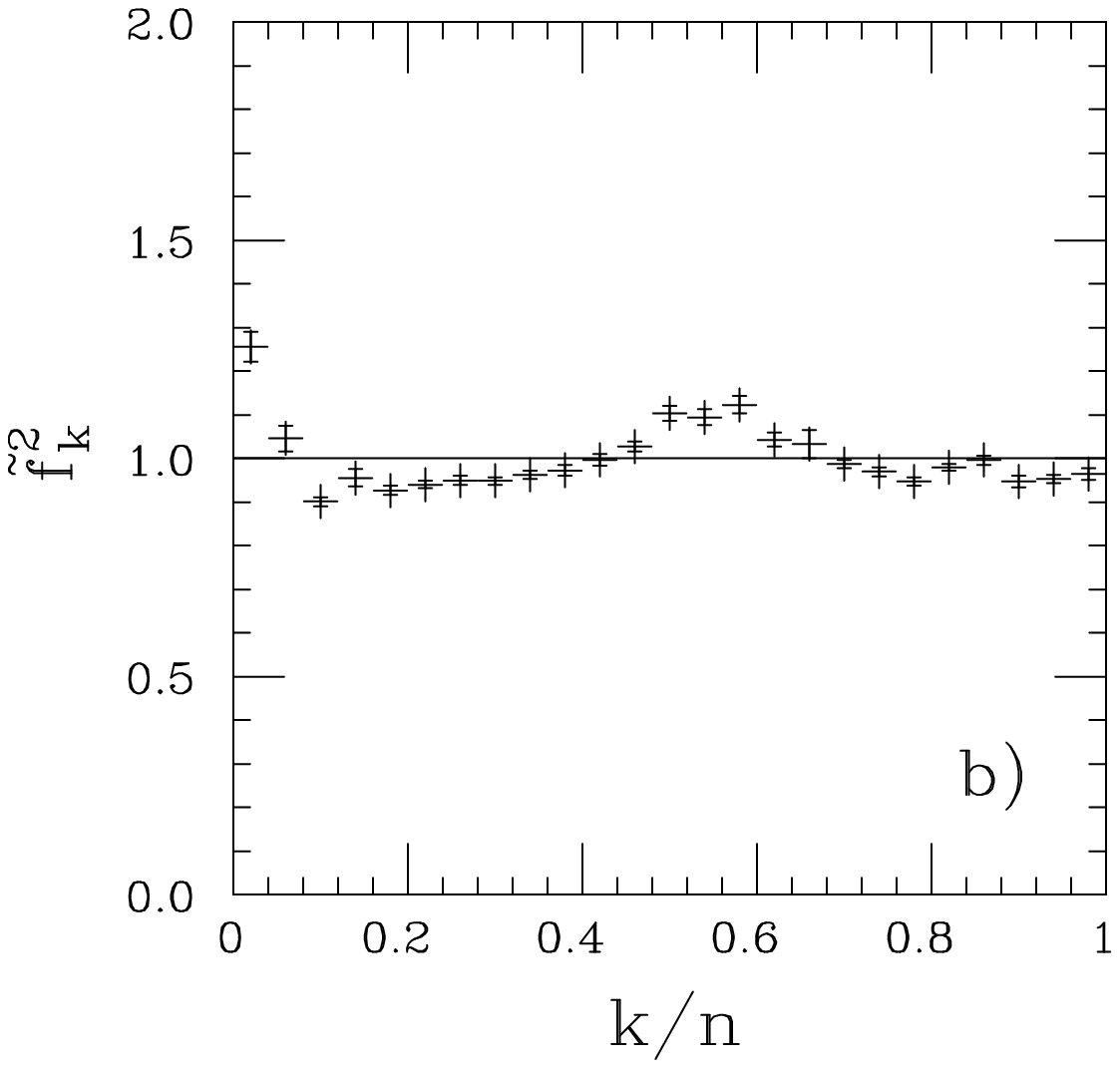,width=10.5cm,height=14cm}}
\vspace{-43mm}
\end{center}
\caption{Normalized squared amplitude $\ft$ against $k/N$ for a) 
$|X|<0.5$ and b) $|X|>3$. The sets of sequences considered are the same
as in Fig.~3.}
\label{fig:5}
\end{figure}

Next we compare the behavior of the Fourier components for small
and large $|X|$. In Fig.~\ref{fig:5} we have plotted the normalized
squared amplitude $\ft$ against $k/N$ for $|X|<0.5$ and $|X|>3$.
Let us first consider the region of small and medium wave-length.
Here the results for the two intervals in $|X|$ are similar. As one
might have expected, there is a peak around the wave-length corresponding 
to $\alpha$-helix structure, $2N/k=3.6$. Away from this peak, 
the results are very close to those for random sequences. 

At large wave-length, on the other hand, the results show a clear 
$|X|$ dependence, and they differ from the results for random sequences 
for both small and large $|X|$. As can be seen from the figure, 
these components are suppressed for small $|X|$ and strong for large $|X|$.

\subsection{Tests on Non-Redundant Sets}

A general problem in the statistical analysis of proteins is the presence of
homologies since these may shift away distributions from an ideal set of 
independent samples.

In order to test for effects due to homologies, we redid the analysis above 
using a set of 486 
selected sequences~\cite{hobohm2}\footnote{The March 1996 edition was used.}
from the Protein Data Bank~\cite{PDB}. 
This set was obtained by allowing for a maximum of 25\% sequence similarity 
for aligned subsequences of more than 80 residues~\cite{hobohm1}.
Within this set of minimally redundant 
sequences, 185 with $|X|<0.5$ and 5 with $|X|>3.0$ 
qualified for analysis. The results for $|X|<0.5$ are within statistics 
identical to  those described above. For  $|X|>3.0$ the results are not in 
conflict with the results above but quantitative comparisons are not 
meaningful due to the extremely small sample size.

The fact that our results survive when limiting ourselves to 
non-redundant proteins, implying a substantial cut in number of proteins 
involved in the 
analysis, makes the evidence of non-randomness even stronger.

\section{A Simplified Synthetic Model}

In this section we carry through the same hydrophobicity analysis as above for 
a simple toy model for proteins~\cite{still1,still2} with binary amino acids 
- The {\bf AB} model. Due to its 
simplicity and the relatively small sizes involved the folding properties 
of this model have been studied to quite some 
detail~\cite{irback,irback1}. The question we want to 
address here is whether the sequences, which have good folding properties in 
the AB model, deviate from the non-folding ones in a way qualitatively similar 
to what was found for the small-$|X|$ functional proteins above. As will be 
shown below this is indeed the case.

\subsection{The AB Model}

In this model, there are two kinds of residues with $\sigma_i=\pm 1$ (A and B) 
respectively. These are linked by rigid bonds to form linear chains in two 
dimensions. The interactions between the residues are given by 
$\sigma_i$-dependent Lennard-Jones potentials such that ($++$) 
is strongly attractive, ($--$) weakly attractive and ($+-$) repulsive. 
In Ref.~\cite{irback} the thermodynamics of this system at 
low temperature was studied using the hybrid Monte Carlo method. Fluctuations 
in the shape for a given chain were studied by measuring the mean-square 
distance $\delta^2$ between pairs of configurations; the probability 
distribution of $\delta^2$, for fixed temperature and sequence, describes the 
magnitude of the thermodynamically relevant fluctuations. It is suggestive 
to interpret a low average $\delta^2$ as a signal for good folding and 
stability properties. Recently an attempt to understand the systematics of 
how low $\delta^2$ values relate to the $\sigma_i$ sequence was 
pursued \cite{irback1}. In this work 300 randomly selected sequences with 14 A 
and 6 B residues were studied, using an improved Monte Carlo 
method \cite{marinari,irback}. The sequences were classified as having good 
folding properties if the average $\delta^2$ was less than 0.3, or if the 
probability of $\delta^2<0.1$ was greater than 0.35. This yielded 
a total of 37 good folders (roughly 10 \%). 

\subsection{Results}

Using the 37 good folding sequences, we have repeated the analysis
of the previous section. This set of sequences is fairly small, but
has the advantage that it is has been generated in a bias-free way.
Statistical errors given in this section have been obtained by taking  
the results for different sequences as independent measurements.     

\begin{figure}[tbp]
\begin{center}
\vspace{-45mm}
\mbox{\hspace{0mm}\psfig{figure=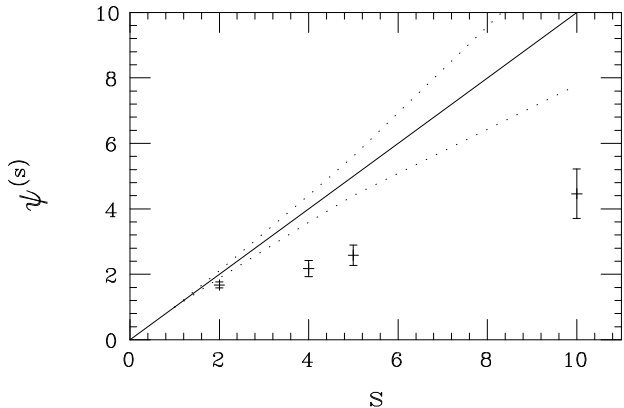,width=10.5cm,height=14cm}}
\vspace{-45mm}
\end{center}
\caption{Mean-square fluctuation of the block variables, $\fl$, against block 
size $s$ for good folding sequences in the AB model. Also shown are
the mean $s$ (full line) and the $s\pm\sigma$ band (bounded by dotted lines) 
for random sequences.}   
\label{fig:6}
\end{figure}

In Fig.~\ref{fig:6} we show the mean-square fluctuation of the block
variables, $\fl$. 
The average of $\fl$ over 37 random sequences has an approximately 
Gaussian distribution, with mean $s$ and a standard deviation 
$\sigma$ that can be obtained by using the results of Appendix A. In the 
figure we have indicated the position of the $s\pm\sigma$ band. We see that 
the data points lie clearly below this band.  

\begin{figure}[tbp]
\begin{center}
\vspace{-45mm}
\mbox{\hspace{0mm}\psfig{figure=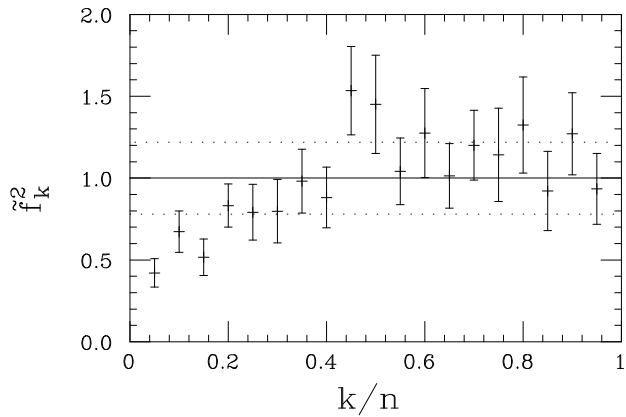,width=10.5cm,height=14cm}}
\vspace{-45mm}
\end{center}
\caption{Normalized squared amplitude $\ft$ against $k/N$ for good folding
sequences in the AB model. The full line and dots are as in Fig.~6.
The standard deviation for random sequences can be obtained by using the
results of Appendix B.} 
\label{fig:7}
\end{figure}

Our results for the squared Fourier amplitude are shown in Fig.~\ref{fig:7}. 
Although the statistical errors on this quantity 
are large, there are clear deviations from the result for random sequences
at large wave-length. We see that components corresponding to 
large wave-lengths are suppressed.  

These results show that good folding sequences in the AB model tend to
exhibit small block fluctuations and weak Fourier components at 
large wave-length. Qualitatively, the results are very similar to those  
obtained in the previous section for small $|X|$.
     
\subsection{Interpretation of the Results}

In this paper we have compared various results with those for random sequences.
Random sequences correspond to a situation in which there is (essentially)
no correlation between $\sigma_i$ and $\sigma_j$ for $i\ne j$. A simple but 
instructive way to introduce non-zero correlations into the system is to 
consider the one-dimensional Ising model. In this model there are $N$ 
``spins'' $\sigma_i$ that take the values $\pm 1$, and each configuration is 
given a statistical weight 
\beq
P\propto\exp\Bigl(K\sum_{i=1}^{N-1}\sigma_i\sigma_{i+1}\Bigr)
\label{ising}
\eeq
As in our previous calculations, we consider configurations with a fixed 
number of positive spins, i.e., the magnetization 
$M=\sum_{i=1}^N\sigma_i=2N_+-N$ is held fixed. Also, as before free boundary  conditions are assumed.

The properties of this system are determined by the parameter $K$. Neighboring
spins tend to point in the same direction if $K>0$ (ferromagnet), and in
opposite directions if $K<0$ (antiferromagnet). For $K=0$ we recover the (random) 
system studied previously.

\begin{figure}[tbp]
\begin{center}
\vspace{-43mm}
\mbox{\hspace{-31mm}\psfig{figure=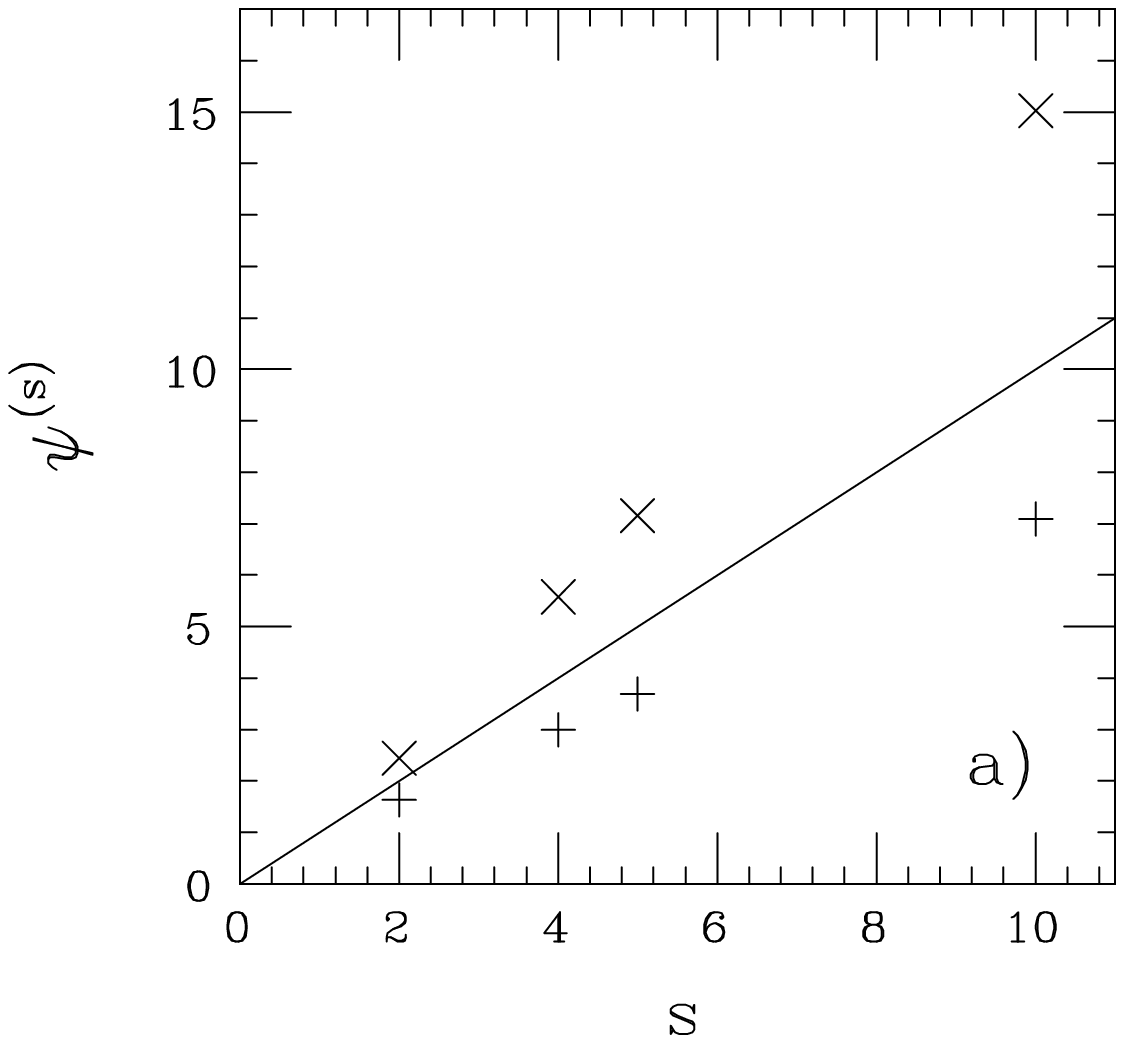,width=10.5cm,height=14cm}
\hspace{-30mm}
\psfig{figure=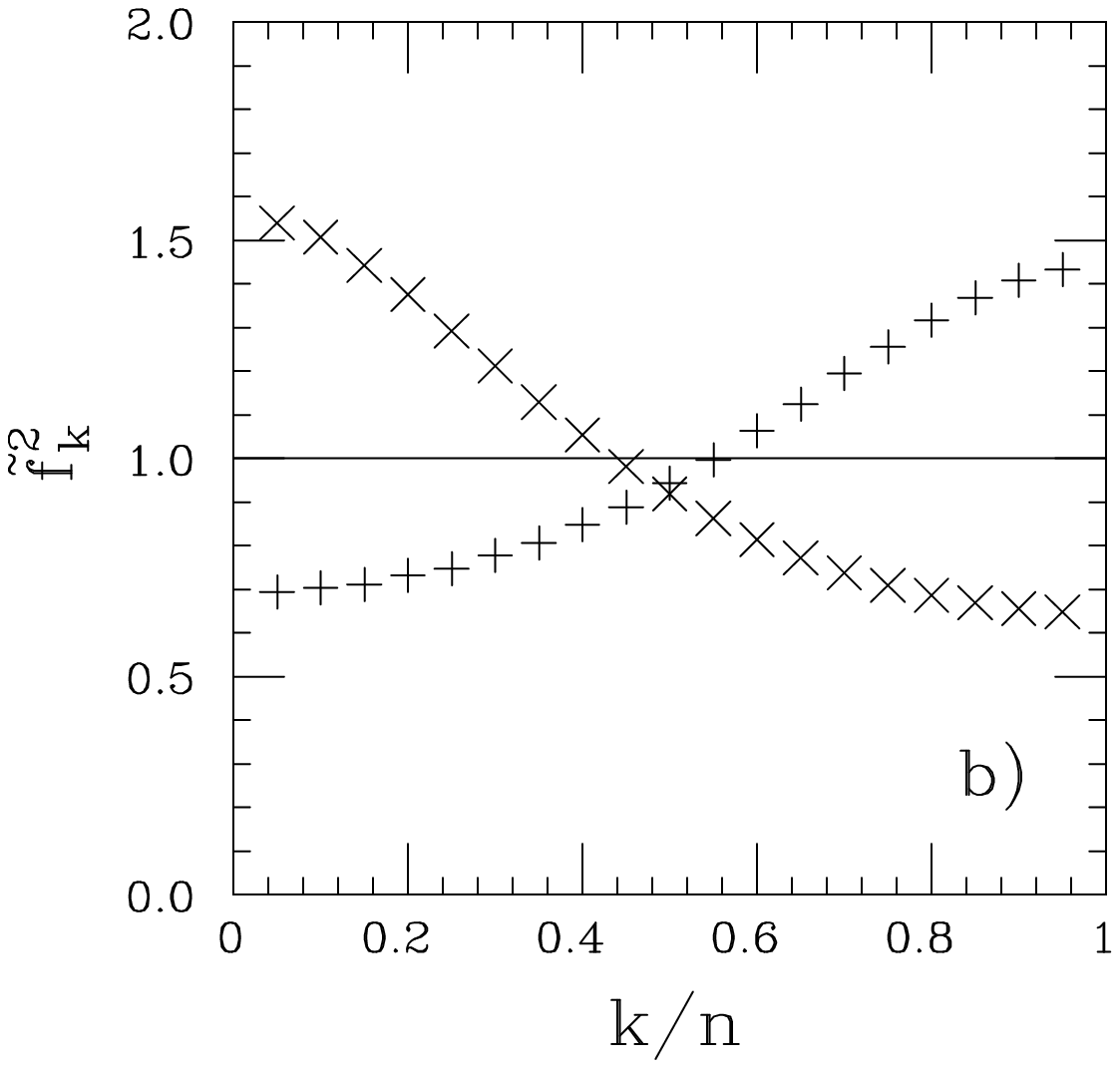,width=10.5cm,height=14cm}
}
\vspace{-43mm}
\end{center}
\caption{a) Mean-square fluctuation of the block variable, $\fl$, and b) normalized
squared amplitude $\ft$ for the Ising model with $K=-0.25$ (+) and 
$K=0.25$ ($\times$). The lines correspond to $K=0$.}
\label{fig:8}
\end{figure}

To illustrate the behavior of the system for non-zero $K$, we show in Fig.~\ref{fig:8} 
results obtained at $K=\pm0.25$. As in our AB calculations, 
we have taken $N=20$ and $N_+=14$. At $K=0.25$ we see that block fluctuations    
are large and that Fourier components with large wave-length are strong, while
the behavior is the opposite at $K=-0.25$. This means that the results for 
this antiferromagnetic system are similar to those obtained for good folding sequences 
in the AB model and for protein sequences with small $|X|$. On the other hand, 
the results for the ferromagnetic system resemble those for proteins with large $|X|$.        

\section{Summary}

We have demonstrated that the statistical distribution of hydrophobic 
residues along chains of functional proteins are non-random. This result is in 
contrast with what was concluded in Ref.~\cite{white}. An important 
reason for this difference is probably that the blocking and Fourier 
analysis methods are able to capture long range correlations more efficiently 
than the method of Ref.~\cite{white}. 
In Ref.~\cite{pande}, on the other hand, a method more similar to ours was
used and deviations from random behavior were observed, but the deviations
may seem to differ in nature from what we have found. However, it
is important to note that these authors focused on hydrophilicity rather
than hydrophobicity, as they used a binary classification in which five
strongly hydrophilic residues formed one group. 
Also, the interpretation of the results of Ref.~\cite{pande} is somewhat
unclear, as no distinction was made between the interior and the ends
of the sequences. When limiting the data set to 
non-redundant protein chains 
the results from the analysis are unaffected. Hence we consider our evidence 
for non-randomness as being quite robust.

We have also applied our analysis method to a toy model data base (AB model) 
where chains with good folding properties were distinguished from the rest. 
The hydrophobicity distributions of the good folding sequences differ from 
random ones in qualitatively the same way as for the low-$|X|$ functional 
protein analysis. It is tempting to interpret this similarity as indicating 
that only those proteins with good folding properties have 
survived the evolution.

The deviation from randomness in the AB model case can be understood as 
originating from anticorrelations among the residues. The effects of 
correlations and anticorrelations on the observables considered were
illustrated by using the simple one-dimensional Ising model. 

Our analysis has been a statistical one in the sense that distributions 
are being compared. Given our encouraging result, it might be possible to 
reach the ultimate goal of being 
able to classify individual sequences in terms of belonging to one category 
or the other. This might be feasable by considering suitable cuts in the block
and Fourier quantities. Very likely, one then needs to augment the method with 
additional discriminative variables and an automated procedure like 
artificial neural networks for setting the cuts.

Our analysis has been confined to binary hydrophobicity assignments.  
The results presented are insensitive to minor modifications of these 
assignments. We do not expect the results to change significantly if instead 
of binary assignments multivalued ones are used.

\newpage

\setcounter{section}{0}
\renewcommand{\thesection}{Appendix \Alph{section}.}
\renewcommand{\theequation}{\Alph{section}\arabic{equation}}
\section{}
\setcounter{equation}{0}

\section*{Variance of $\fl$}

In this Appendix we give the variance of $\fl$ (see Eq.~\ref{msfluct}). 
The average of $\fl$ over all sequences with fixed composition, $N_+$ and 
$N_- =N-N_+$,  is given by
\beq
\evf{\fl}={1\over K}\sum_{i,j=1}^s c_{ij}
\eeq
where $c_{ij}$ is the connected correlation between $\sigma_i$ and 
$\sigma_j$, for which one finds 
\beq
c_{ij}=\evf{\sigma_i\sigma_j}-\evf{\sigma_i}\evf{\sigma_j}=
                             \left\{ \begin{array}{ll}
                               {4N_+N_-\over N^2}     & \mbox{if $i=j$}\\
                               &\\
                              -{4N_+N -\over N^2(N-1)} & \mbox{if $i\ne j$}
                              \end{array} \right.\\
\eeq
Using this, one obtains $\evf{\fl}=s$ (Eq.~\ref{lin}).
The off-diagonal correlation, $c_{ij}$ with $i\ne j$, has to be negative
since $\sum_{i=1}^N\sum_{j=1}^Nc_{ij}=0$, but vanishes in
the limit $N\to\infty$. 

The variance can be computed in a similar way. In addition to $c_{ij}$, one
then needs the correlation between four $\sigma_i$'s. One finds  
\beq
\evf{\fl{} ^2}-\evf{\fl}^2=2s^2(s-1)N^{-1}K^{-2}G(s,N,N_+)
\eeq
where
\begin{eqnarray}
G(s,N,N_+) &=&1+2(s-2){\de^2-N\over N(N-1)}\\ \nonumber
&  &- (2s-3){\de^4-6N\de^2+3N^2+8\de^2-6N\over N(N-1)(N-2)(N-3)} \\ \nonumber
& &+ {1\over 2}(s-1)\cdot\Bigl({(4N-6)\de^4\over N(N-1)^2(N-2)(N-3)} \\  \nonumber
& &+ {-6N\de^2+3N^2+8\de^2-6N\over (N-1)(N-2)(N-3)}+{2\de^2-N\over (N-1)^2} \Bigr) \nonumber
\end{eqnarray}

In the limit $N\to\infty$, for fixed $s$, this expression simplifies to 
Eq.~\ref{varmsfluct}.

\newpage

\section{}
\setcounter{equation}{0}

\section*{Fourier Transforms of Random Walk Representations}

In this Appendix  Fourier transform moments of  random walk representations 
are listed. The expressions are more general than what is required for 
binary hydrophobicity assignments. In order to do this we first list the following 
basic quantities for sequences of length $N$:

\begin{itemize}
\item Moment of order $k$, $m_k$. 
\subitem $m_k={1\over N}\sum_{i=1}^{N}\sigma_i^k$.

\item Cumulant of order $k$, $c_k$. 
\subitem $c_2=m_2-m_1^2$
\subitem $c_4=m_4-4m_3m_1-3m_2^2+12m_2m_1^2-6m_1^4$

\item Random walk, $\rho_n$. 
\subitem $\rho_0=0$
\subitem $\rho _n=\sum_{i=1}^n \sigma_i-nm_1\qquad i=1,\ldots,N\qquad 
(\rho _N=0)$
 
\item Sine transform of $\rho_n$, $f_k$
\subitem $f_k=\sum_{n=1}^{N-1}\rho _n\sin{\pi kn\over N}\qquad k=1,\ldots,N-1$ 

\end{itemize}

Averaging over all sequences with fixed composition,  
i.e., all permutations of  $\sigma_i$, one obtains
\bea
\ev{\rho_n}&=&0\\
\ev{\rho_n^2}&=&{N^2c_2\over N-1}\cdot{n\over N}(1-{n\over N})\\
\ev{f_k}&=&0\\
\ev{f_k^2}&=&{N^2c_2\over 2(N-1)}\cdot {1\over (2\sin{\pi k\over 2N})^2}\\
\ev{f_k^4}&=&{3N^4\over 4(N-1)(N-2)}\Bigl[c_2^2-{1\over 2N}(c_4+6c_2^2)
\nonumber\\
&&-\delta_{2k,N}{1\over 24(N-3)}\Bigl(c_4+{1\over N}(c_4+6c_2^2)\Bigr)\Bigr]
\cdot {1\over (2\sin{\pi k\over 2N})^4}
\eea
For the binary scale $\sigma_i=\pm 1$ one has $c_2=4N_+(N-N_+)/N^2$ and
Eq.~B4 becomes Eq.~\ref{fk2}.

\newpage

\section{}
\setcounter{equation}{0}

\section*{Removal of Uncertain Sequences}

In our analysis we have removed "uncertain sequences"  from the SWISS-PROT 
database by ignoring  all entries containing  the following feature 
keys in their feature key table:
\begin{itemize}
 \item  {\bf UNSURE} - Indicates that there are uncertainties in the sequence.
 \item  {\bf NON\_CONS } - Indicates that two residues in a sequence are 
not  consecutive and that there are a number of unsequenced residues in between.
 \item  {\bf NON\_TER } - The residue at an extremity of the sequence is not 
the terminal residue.
\end{itemize}
This reduces the size of the SWISS-PROT database from 43470 to 38050 protein entries. 

Furthermore, when analyzing the interior parts of protein sequences,  
sequences containing the following letters 
\begin{itemize}
 \item {\bf B}  denoting Aspartic acid or Asparagine.
\item {\bf Z}  denoting Glutamine or Glutamic acid.
 \item {\bf X}  denoting any amino acid.
\end{itemize}
within the interior are removed.


\newpage

\begin{thebibliography}{99}

\newcommand	{\B}     {{\it Biopolymers\ }}
\newcommand	{\BC}    {{\it Biophys.\ Chem.\ }}
\newcommand     {\BJ}    {{\it Biophys.\ J.\ }}
\newcommand	{\EL}	 {{\it Europhys.\ Lett.\ }}
\newcommand     {\JCC}   {{\it J.\ Comp.\ Chem.\ }}
\newcommand	{\JCoP}	 {{\it J.\ Comp.\ Phys.\ }}
\newcommand	{\JCP}	 {{\it J.\ Chem.\ Phys.\ }}
\newcommand	{\JMB}	 {{\it J.\ Mol.\ Biol.\ }}
\newcommand	{\JP}	 {{\it J.\ Phys.\ }}
\newcommand     {\JPC}   {{\it J.\ Phys.\ Chem.\ }}
\newcommand	{\JSP}	 {{\it J.\ Stat.\ Phys.\ }}
\newcommand     {\M}     {{\it Macromolecules\ }}
\newcommand     {\MC}    {{\it Makromol.\ Chem.,\ Theory Simul.\ }}
\newcommand     {\MP}    {{\it Molec.\ Phys.\ }}
\newcommand     {\NAR}   {{\it Nucleic\ Acids\ Res.\ }}
\newcommand     {\Prot}  {{\it Proteins\ }}
\newcommand     {\Pa}    {{\it Physica\ }}
\newcommand	{\PL} 	 {{\it Phys.\ Lett.\ }}
\newcommand	{\PNAS}  {{\it Proc.\ Natl.\ Acad.\ Sci.\ USA\ }}
\newcommand	{\PR}	 {{\it Phys.\ Rev.\ }}
\newcommand	{\PRL}	 {{\it Phys.\ Rev.\ Lett.\ }}
\newcommand	{\ZP}	 {{\it Z.\ Physik\ }}
\newcommand     {\PROTSC}{{\it Protein\ Science\ }}

\bibitem{white} White, S. H. \& Jacobs, R. E. (1990) 
\BJ {\bf 57}, 911-921. 

\bibitem{shakhnovich} Shakhnovich, E. I. \& Gutin, A. M. (1993) 
\PNAS {\bf 90}, 7195-7199.

\bibitem{pande} Pande, V. S., Grosberg, A. Y. \& Tanaka, T. (1994)
\PNAS {\bf 91}, 12972-12975.
 
\bibitem{sali} \u{S}ali, A., Shakhnovich, E. \& Karplus, M. (1994)
\JMB {\bf 235}, 1614-1636. 

\bibitem{irback} Irb\"ack, A. \&  Potthast, F. (1995)
\JCP {\bf 103},  10298-10305.

\bibitem{swiss-prot} Bairoch, A. \& Boeckmann, B. (1994)
\NAR {\bf 22}, 3578-3580.

\bibitem{still1} Stillinger, F. H., Head-Gordon, T. \& Hirschfeld, C. L. (1993)
\PR {\bf E48}, 1469-1477.

\bibitem{still2} Head-Gordon, T. \& Stillinger, F. H. (1993)
\PR {\bf E48}, 1502-1515.

\bibitem{irback1}  Irb\"ack, A., Peterson, C. \& Potthast, F. preprint LU TP 96-12, chem-ph/9605002
(submitted to \PR {\bf E}).

\bibitem{marinari}
Marinari, E. \& Parisi, G. (1992)
\EL {\bf 19}, 451-458.

\bibitem{eisenberg} Eisenberg, D., Weiss, R. M. \& Terwilliger, T. C. (1984)
\PNAS {\bf 81}, 140-144.

\bibitem{hobohm2} Hobohm, U. \& Sander, C. (1994)
\PROTSC {\bf 3}, 522-524.

\bibitem{PDB} Bernstein, F. C., Koetzle, T. F., Williams, G. J. B., Meyer, E. F., Brice, M. D.,
Rodgers, J. R., Kennard, O., Shimanouchi, T. \& Tasumi, M. (1977)
\JMB {\bf112}, 535-542.

\bibitem{hobohm1} Hobohm, U.,  Scharf, M., Schneider, R. \& Sander, C. (1992)
\PROTSC {\bf 1}, 409-417.


\end{thebibliography}
\end{document}